\documentclass[twocolumn,showpacs,preprintnumbers,amsmath,amssymb]{revtex4}
\usepackage{graphicx}% Include figure files
\usepackage{dcolumn}% Align table columns on decimal point
\usepackage{bm}% bold math

\def\ii{{\mathrm{i}}}

\def\dd{{\mathrm{d}}}

\def\bracket#1{\langle #1 \rangle}

\def\sub#1{_{\mathrm{#1}}}
\def\up#1{^{\mathrm{#1}}}
\def\Vec#1{\mbox{\boldmath $#1$}}

\begin{document}

\preprint{APS/123-QED}

\title{Inertial Range and the Kolmogorov Spectrum of Quantum Turbulence}
% Force line breaks with \\

\author{Makoto Tsubota}
\author{Michikazu Kobayashi}%
\affiliation{Faculty of Science, Osaka City University, Sugimoto3-3-138, Sumiyoshi-ku, Osaka558-8585, Japan}%

%\affiliation{

\date{\today}% It is always \today, today,

\begin{abstract}
Quantum turbulence is numerically studied by solving the Gross-Pitaevskii equation. Introducing both the energy dissipation at small scales and the energy injection at large scales, we succeed in obtaining the steady turbulence made by the balance of the injection and the dissipation of the energy. The inertial range takes the Kolmogorov spectrum and the energy flux is consistent with the energy dissipation rate at small scales. These results confirm the properties of the inertial range of quantum turbulence, proposing a prototype of turbulence much simpler than conventional classical turbulence.
\end{abstract}

\maketitle

Quantum turbulence (QT) consisting of quantized vortices has become a great subject not only for investigating the issues of superfluidity in the proper field of low temperature physics but also for studying a prototype much simpler than conventional classical turbulence (CT).  The steady state of fully developed CT of an incompressible fluid takes the characteristic statistical law for the energy spectrum \cite{Frisch}. The energy,  injected into the fluid at some large scales in the energy-containing range,  is transferred in the inertial range to smaller scales without been dissipated.  In the inertial range the energy spectrum takes the Kolmogorov law
\begin{equation}
E(k)=C\varepsilon^{2/3}k^{-5/3}.\label{eq-Kolmogorov}
\end{equation}
Here the energy spectrum $E(k)$ is defined as $E=\int\dd k\:E(k)$, where $E$ is the kinetic energy per unit mass and $k$ is the wave number from the Fourier transformation of the velocity field. The energy transferred to the energy-dissipative range is dissipated by the viscosity with the dissipation rate $\varepsilon$ of Eq. (\ref{eq-Kolmogorov}), which is equal to the energy flux $\Pi$ in the inertial range. The Kolmogorov constant $C$ is a dimensionless parameter of order unity.

The inertial range is believed to be sustained by the self-similar Richardson cascade process, in which large eddies are broken up to smaller eddies through reconnections. However, the Richardson cascade has never been confirmed clearly in CT, because it is impossible to definitely identify each eddy. In contrast, quantized vortices are definite and stable topological defects with the quantized circulation, consequently QT may give us a prototype of turbulence much simpler than CT \cite{Kobayashi}.  Some experimental \cite{Maurer, Stalp} or theoretical works \cite{Nore, Araki} studied the Kolmogorov law for QT, but it was very difficult to control the energy dissipation rate $\varepsilon$.

Recently we made the numerical simulation of the Gross-Pitaevskii (GP) equation which describes a quantum fluid at zero temperature \cite{Kobayashi}. Here we introduced a dissipative term that works only in the smaller scales than the healing length, in order to remove compressible short-wavelength excitations created through vortex reconnections which may hinder the cascade process.  We thus obtained the Kolmogorov spectrum for decaying turbulence starting from the random configuration of the velocity field. In this work, we also inject energy into the fluid  at large scales by moving the random potential, thus obtaining steady turbulence sustained by the balance of the injection and dissipation of the energy. The energy spectrum of the incompressible kinetic energy still takes the Kolmogorov law, and the energy flux $\Pi$ is almost independent of the scales in the inertial range and consistent with the energy dissipation rate in the energy-dissipative range. These results make clear the properties of the inertial range of QT.

\section{The model of the GP equation for steady turbulence.} \label{sec-model}
We study numerically the Fourier transformed GP equation
\begin{widetext}
\begin{equation}
  [\ii-\gamma(k)]\frac{\partial}{\partial t}\tilde{\Phi}(\Vec{k},t)=[k^2-\mu(t)]\tilde{\Phi}(\Vec{k},t)+\frac{g}{L^6}\sum_{\Vec{k}_1,\Vec{k}_2}\tilde{\Phi}(\Vec{k}_1,t)\tilde{\Phi}^\ast(\Vec{k}_2,t)\tilde{\Phi}(\Vec{k}-\Vec{k}_1+\Vec{k}_2,t)+\frac{1}{L^3}\sum_{\Vec{k}_1}\tilde{V}(\Vec{k}_1,t)\tilde{\Phi}(\Vec{k}-\Vec{k}_1,t).\label{eq-Fourier-GP}
\end{equation}
\end{widetext}
Here $\tilde{\Phi}(\Vec{k},t)$ is the Fourier transformation of the macroscopic wave function $\Phi(\Vec{x},t)=f(\Vec{x},t)\exp[\ii\phi(\Vec{x},t)]$, $g$ is the coupling constant, $L$ the system size.  The dissipation is given by  
\begin{equation} \gamma(k)=\gamma_0\theta(k-2\pi/\xi), \end{equation} which works only in the scales smaller than the healing length $\xi=1/f\sqrt{g}$ to remove compressible short-wavelength excitations.  The chemical potential $\mu(t)$ depends on time to conserve the total number of particles $N=\int\dd\Vec{x}\:f(\Vec{x},t)^2$.  The Fourier transformation of the moving random potential $V(\Vec{x},t)$ is $\tilde{V}(\Vec{k},t)$, which moves so that it may satisfy the two-point correlation
\begin{equation}
  \bracket{V(\Vec{x},t)V(\Vec{x}^{\prime},t^{\prime})}=V_0^2\exp\Big[-\frac{(\Vec{x}-\Vec{x}^{\prime})^2}{2X_0^2}-\frac{(t-t^{\prime})^2}{2T_0^2}\Big],\label{eq-random-potential}
\end{equation}
where $V_0$, $X_0$ and $T_0$ are the characteristic strength and scales of space and time, respectively. Thus $\tilde{V}(\Vec{k},t)$ injectes the energy to the fluid at the scale $X_0$ to set the energy-containing range. 

We numerically solve Eq. (\ref{eq-Fourier-GP}) in a periodic box with spatial resolution containing $256^3$ grid points. We consider the case of $g=1$. As numerical parameters, we use a spatial resolution $\Delta x=0.125$ and $L=32$, where the length scale is normalized by the healing length $\xi$. The parameters of the moving random potential are $V_0=50$, $X_0=4$ and $T_0=6.4\times 10^{-2}$. The numerical integral is made by the pseudo-spectral method in space and the Runge-Kutta-Verner method in time with a time resolution $\Delta t=1\times 10^{-4}$. We start from an initial configuration in which both the density and the phase of the wave function are uniform.  The system gets to steady turbulence after $t \simeq 25$ when both the total energy $E$ and the incompressible kinetic energy $E\sub{kin}\up{i}$, given by
\begin{subequations}
\begin{align} E=&\frac1N \int\dd\Vec{x}\:\Phi^{\ast}[\nabla^2+\frac{g}{2} f^2]\Phi,\label{eq-energy}\\
E\sub{kin}\up{i}=&\frac1N \int\dd\Vec{x}\:[\{f\nabla\phi\}\up{i}]^2,\label{eq-incompressible-energy}
\end{align}
\end{subequations}
are almost constant in time. Here $\{\cdots\}$ denotes the incompressible part $\nabla\cdot\{\cdots\}=0$ of the vector fields.  In the steady state we can obtain the dissipation rate $\varepsilon$ from the time derivative of the incompressible kinetic energy $\partial E\sub{kin}\up{i}/\partial t$ after switching off the moving random potential. 

\begin{figure}[btp]
\begin{minipage}{0.49\linewidth}
\begin{center}
\includegraphics[width=0.97\linewidth]{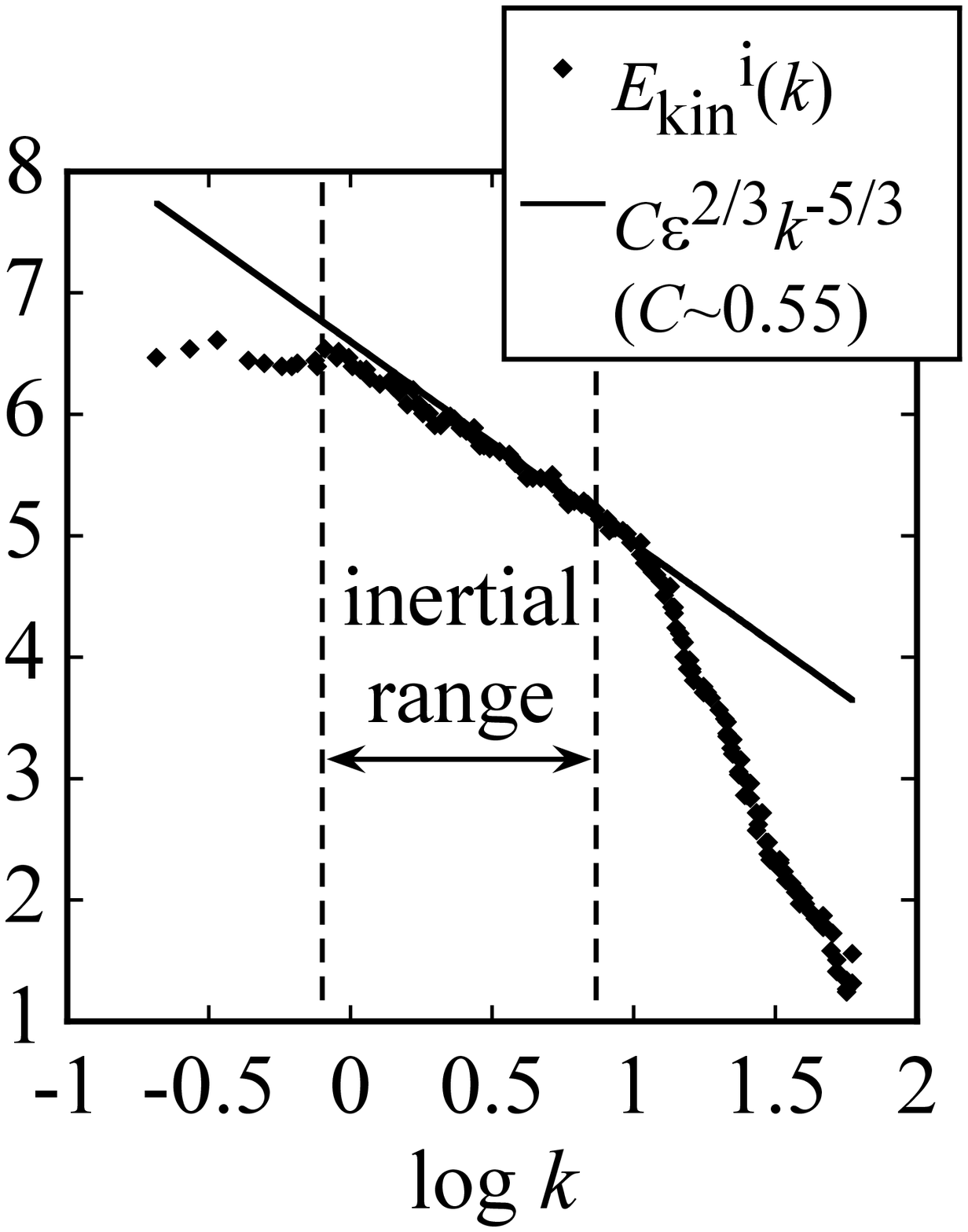}\\
(a)
\end{center}
\end{minipage}
\begin{minipage}{0.49\linewidth}
\begin{center}
\includegraphics[width=0.97\linewidth]{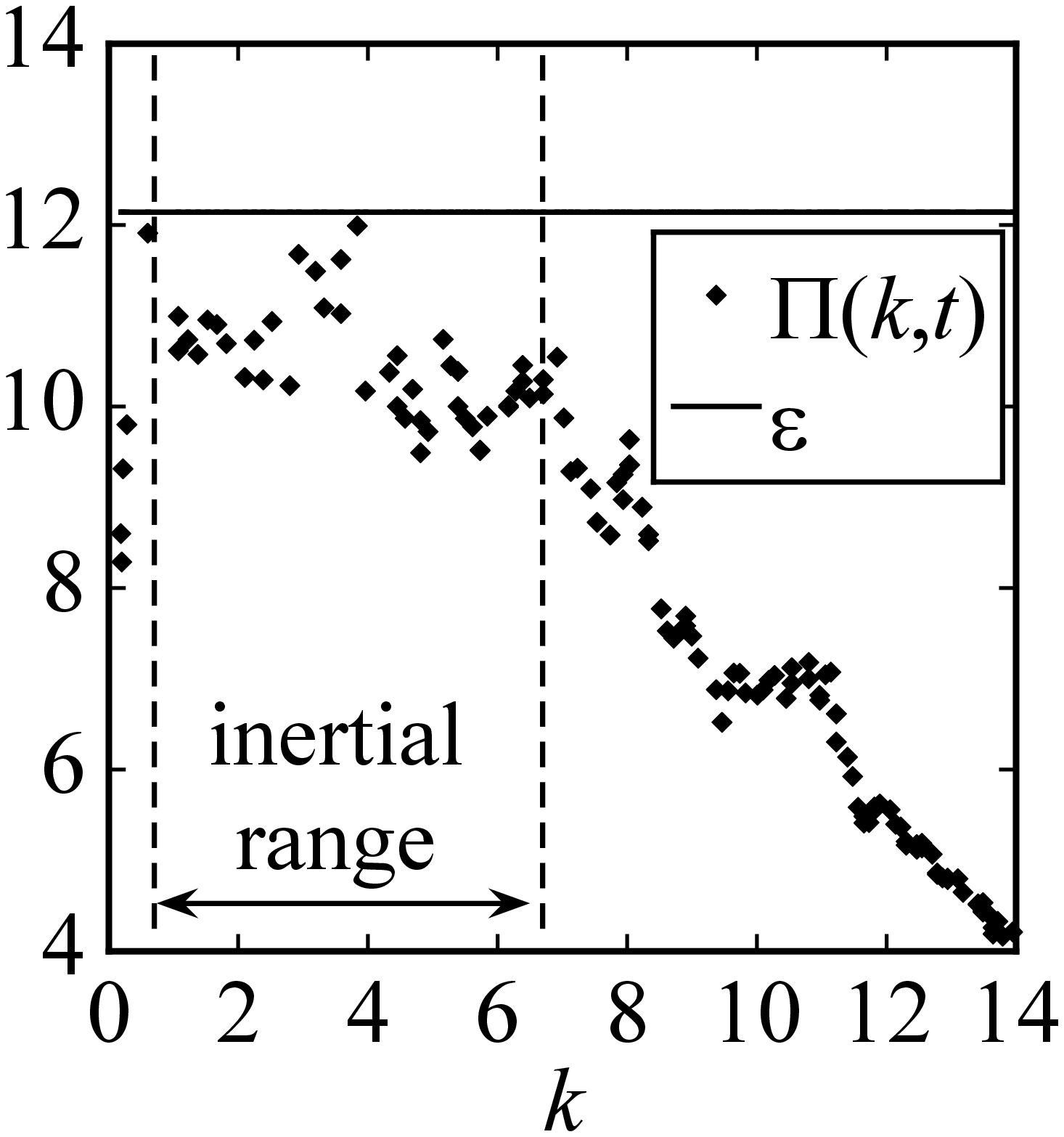}\\
(b)
\end{center}
\end{minipage}
\caption{\label{fig-steady} Dependence of the energy spectrum $E\sub{kin}\up{i}(k)$ (a) and the energy flux $\Pi(k,t)$ (b) of the incompressible kinetic energy on the wave number $k$. Solid lines refer to the Kolmogorov law $C\varepsilon^{2/3}k^{-5/3}$ in (a) and the energy dissipation rate $\varepsilon$ in (b), respectively.}
\end{figure}

In this steady turbulence, we can clearly define the energy-containing range $k<2\pi/X_0$ in which the system takes the energy from the moving random potential, the energy-dissipative range $k>2\pi/\xi$ in which the energy are dissipated by $\gamma(k)$, and the inertial range $2\pi/X_0<k<2\pi/\xi$ between these two ranges. Figure \ref{fig-steady} (a) shows the energy spectrum of the incompressible kinetic energy $E\sub{kin}\up{i}(k)$ defined as $E\sub{kin}\up{i}=\int\dd k\:E\sub{kin}\up{i}(k)$. In the inertial range, the numerically obtained spectrum is quantitatively consistent with the Kolmogorov law using the above obtained $\varepsilon$.
Figure \ref{fig-steady} (b) shows the flux $\Pi(k)$ of the incompressible kinetic energy which is given by the energy budget equation of Eq. (\ref{eq-Fourier-GP}). The energy flux $\Pi(k)$ is about independent of $k$ in the inertial range and close to the energy dissipation rate $\varepsilon$. This result supports strongly the scenario of QT played by quantized vortices; the energy-containing range, the inertial range  and the energy-dissipating range are, respectively, made by their nucleation, their Richardson cascade and their decay to compressible excitations, as believed in eddies of CT.

The detailed studies of this issue are in progress, being reported shortly elsewhere.

\end{document}